\newif\ifdraft
\draftfalse

\newif\ifarxiv
\arxivtrue

\newif\ifeasychair
\easychairtrue

\ifdraft
\documentclass[draft]{llncs}
\else
\documentclass{llncs}
\fi

\newif\iffinal
\finaltrue

\usepackage[T1]{fontenc}
\usepackage[utf8]{inputenc}
\usepackage{lmodern}
\usepackage[scaled=.8]{beramono}
\usepackage{latexsym}
\usepackage{textcomp}
\ifdraft\else
\usepackage{microtype}
\fi

\usepackage[british]{babel}
\usepackage[inline]{enumitem}
\usepackage[babel]{csquotes}
\MakeAutoQuote{“}{”}
\MakeAutoQuote*{‘}{’}
\usepackage{soul}
\setul{.25ex}{}

\usepackage[obeyDraft,textsize=tiny]{todonotes}

\usepackage[backend=bibtex,hyperref=auto,
firstinits=true,style=numeric,
urldate=iso8601,isbn=false,doi=false]{biblatex}
\ifeasychair
\else
\fi
\DeclareNameAlias{author}{last-first/first-last}
\DeclareNameAlias{editor}{last-first/first-last}
\DeclareNameAlias{translator}{last-first/first-last}
\DeclareFieldFormat{labelnumberwidth}{#1.}

\ifarxiv
\newcommand{\href}[2]{#2}
\else
\usepackage{hyperref}
\fi

\begin{document}

\title{The ForMaRE Project –\newline Formal Mathematical Reasoning in Economics\iffinal\thanks{This work has been supported by EPSRC grant EP/J007498/1.  The final publication is available at \texttt{http://link.springer.com}.}\fi}
\author{Christoph Lange\inst{1}
\and Colin Rowat\inst{2}
\and Manfred Kerber\inst{1}}
\institute{%
Computer Science, University of Birmingham, UK%
\and Economics, University of Birmingham, UK\\
Project homepage: \url{http://cs.bham.ac.uk/research/projects/formare/}
}

\maketitle

\begin{abstract}
  The ForMaRE project applies \ul{for}mal \ul{ma}thematical \ul{r}easoning to \ul{e}conomics.  We seek to increase confidence in economics' theoretical results, to aid in discovering new results, and to foster interest in formal methods, i.e.\ computer-aided reasoning, within economics.  To formal methods, we seek to contribute \todo{CL: FYI this aspect is new (from other CICM paper)}user experience feedback from new audiences, as well as new challenge problems.  In the first project year, we continued earlier game theory studies but then focused on auctions, where we are building a toolbox of formalisations, and have started to study matching and financial risk.  
  In parallel to conducting \emph{research} that connects economics and formal methods, we organise events and provide infrastructure to connect both \emph{communities}, from fostering mutual awareness to targeted matchmaking.  These efforts extend beyond economics, towards generally enabling domain experts to use mechanised reasoning.
\end{abstract}

\section{Motivation of the ForMaRE Project}
\label{sec:motivation}

The ForMaRE project applies \ul{for}mal \ul{ma}thematical \ul{r}easoning to \ul{e}conomics.  Theoretical economics \todo{CL: FYI dropped “may be regarded as a branch of applied mathematics”: should be obvious from context}draws on a wide range of mathematics to explore and prove properties of stylised economic environments.  Mathematical formalisation and computer-aided reasoning have been applied there before, most prominently to social choice theory (cf., e.g., \cite{ge-en-11}) and game theory (cf., e.g., \cite{ta-li-11}).  Immediately preceding ForMaRE, we have ourselves formalised pillage games, a particular form of cooperative games, and motivated this as follows at CICM 2011~\cite{ke-ro-ww-11}\todo{CL: slightly shorter phrasing follows; please check whether it's still correct!}:
\begin{enumerate*}
  \item Economics, and particularly cooperative game theory, is a relatively new area for mechanised reasoning (still in 2013) and therefore presents a new set of canonical examples and challenge problems.
  \item Economics typically involves new mathematics in that axioms particular to economics are postulated.  One of the intriguing aspects of cooperative game theory is that, while the mathematical concepts involved are often intelligible to even undergraduate mathematicians, general theories are elusive.  This has made pillage games more amenable to formalisation than research level mathematics.
  \item In economics, as in any other mathematical discipline, establishing new results is an error-prone process, even for Nobel laureates (cf.\ \cite{ke-ro-ww-11} for concrete examples).  As one easily assumes false theorems or overlooks cases in proofs, formalisation and automated validation may increase confidence in results.  Knowledge management facilities provided by mechanised reasoning systems may additionally help to reuse proof efforts and to explore theories to discover new results.
\end{enumerate*}
Despite these potential benefits, economics has so far been formalised almost exclusively by computer scientists, not by economists.

\section{The ForMaRE Strategy}
\label{sec:strategy}

The ForMaRE project, kicked off by the authors in May 2012 and further advised by more than a dozen of external computer scientist and economist collaborators, 
seeks to foster interest in formal methods \emph{within} economics.  Our strategy consists in using this technology to establish new results, building trust in formalisation technology and enabling economists to use it themselves.

\subsection{Establishing New Results}
\label{sec:establ-new-results}

In preparing one of our first activities, an overview of mechanised reasoning for economists (cf.\ sec.~\ref{sec:conn-comp-science}), we realised that exciting work was being done in areas with broader audiences than cooperative games.  We therefore chose to study auctions, matching markets and financial risk.  We have not yet established new results but \todo{CL: FYI after AISB we can afford to be more specific than just saying ``actively exchanging ideas''}have defined first research goals with experts in these fields, some of whose works we cite in the following: \textbf{auctions} are
widely used 
for allocating goods and services.  \todo{CL@CR: OK to keep this example for auctions?  (It's something that gives the CICM audience a concrete imagination.)  Our actual current focus is on case checking, but that's not the main issue with the ICANN auctions, is it?}Novel auctions are constantly being designed -- e.g.\ for allocating new top-level Internet domains~\cite{CA:ApplicantAuction12} -- but their complexity makes it difficult to establish basic properties, including their efficiency \todo{CL@CR: Their efficiency is not given as trivially as in Vickrey, but is it hard to find out?} i.e.\ give a domain to the registrar who values it highest and is therefore expected to utilise it best.  \todo{CL@CR/@MK: any need to revise this (e.g.\ be more specific) after AISB?}\textbf{Matching} problems occur, e.g., in health care (matching kidney donors to patients) and in education (children to schools)~\cite{SonmezUnver:Matching:11}.  Impossibility results are of particular interest here; they rely on finding rich counter examples.  \todo{CL@CR/@MK: any need to revise this (e.g.\ be more specific) after AISB?}Finally, modern \textbf{finance} relies on models to price assets or to compute risk, but banks and regulators still validate and check such models manually.  One research challenge is to develop minimal test portfolios that ensure that capital models incorporate relevant risk factors~\cite{Vosloo:ModelValidTestPortfFinancReg:13}.

\subsection{Building Trust in Formalisation Technology}
\label{sec:build-trust-form}

Economic theorists typically have a solid mathematics background.  There is a field `computational economics'; however, it is mainly concerned with \emph{numerical} computation of solutions or simulations~\cite{InitCompEcon}.  Contemporary economists still prove their theorems using pen and paper.  While we aim at establishing new results to showcase the potential of formal methods (see above), we also seek to establish confidence in formal methods within the economics community.  Thus, as a first step, we have demonstrated the reliability of formal methods by \emph{re-establishing known results}.  Computer scientists have previously done so by formalising some of the many known proofs of Arrow's impossibility theorem, a central result of social choice theory~\cite{nip-09,wie-09}.  We have started to formalise the review of an influential auction theory textbook~\cite{mas-04} in four theorem proving systems in parallel, collaborating with their developers or expert users~\cite{LangeEtAl:CompProvAuctThy13}.  This formalisation, currently covering \todo{CL: TODO ask MC about progress with proposition 2}Vickrey's theorem on second price auctions, constitutes the core of an Auction Theory Toolbox (ATT~\cite{AuctionTheoryToolbox}).  The review covers 12 more canonical results for single good auctions\todo{CL@CR: FYI in the authors' response we were more specific (including theorems on revenue equivalence across auction types, as well as consequences of violating the assumptions underlying the equivalence results), but there's no space for this.}.  We plan to extend the ATT, including new auction designs as well, and welcome contributions from the community.\todo{CL: FYI not addressing Rev.\ 1 request to “perhaps mention some other known results the authors plan to tackle, e.g. in risk management?” – no space, and not really concrete yet.}\todo{CL: FYI mentioning trusted proofs as future work would make sense, but we have nothing to show \emph{so far}, and there's more space for that in the other paper.}

\subsection{Enabling Economists to use Mechanised Reasoning}
\label{sec:enabl-econ-use}

Ultimately we aim at enabling economists to formalise their own designs and validate them themselves, or at least \todo{CL: OK to put it this way?  Let's be open here and neither commit to enabling only economists, nor to always assuming presence of formalisation specialists.}to train specialists beyond the core mechanised reasoning community, who will assist economists with formalisation – \todo{CL: OK?}just like lawyers assist with legal issues.  For users without a strong mechanised reasoning background the complexity and abundance of formalised languages and proof assistants poses an adoption barrier.  In selected fields, we will provide toolboxes of ready-to-use formalisations of basic concepts, including definitions and essential properties, and guides to extending and applying these toolboxes.  Concretely, this means:
\begin{enumerate*}
\item identifying languages that are
  \begin{enumerate*}
  \item sufficiently expressive while still exhibiting efficient reasoning tasks, that are
  \item learnable for people used to informal textbook notation, and that
  \item have rich libraries of mathematical foundations
  \end{enumerate*}, and
\item identifying proof assistants that
  \begin{enumerate*}
  \item assist with formalisation in a cost-effective way,
  \item facilitate reuse from the toolbox,
  \item whose output is sufficiently comprehensible to help non-experts understand, e.g., why a proof attempt failed, 
and
  \item whose community is supportive towards non-experts.
  \end{enumerate*}
\end{enumerate*}  In building the ATT, we are comparing four different systems, whose philosophies cover a large subset of the spectrum: \todo{CL: Now that the other paper got accepted I dropped all references to systems to save space – OK?}Isabelle (interactive theorem prover, HOL, accessible via a document-oriented IDE
), CASL/Hets (uniform GUI frontend to a wide range of automated FOL provers
), Theorema (automated but configurable theorem prover, HOL with custom FOL and set theory inference rules, Mathematica notebook interface with a textbook-like notation
), and Mizar (automated proof checker, FOL plus set theory
).  For details on these systems and how well they satisfy the requirements, see~\cite{LangeEtAl:CompProvAuctThy13}.

\section{Building, Connecting, and Serving Communities}
\label{sec:community}

In parallel to our research on connecting economics and formal methods, we are conducting community building efforts.

\subsection{Connecting Computer Science and Economics}
\label{sec:conn-comp-science}

With this CICM paper, with an invited lecture at the British Automated Reasoning Workshop~\cite{Kerber:AutoReasEcon13}, and an upcoming tutorial at the German annual computer science meeting themed “computer science adapted to humans, organisation and the environment”~\cite{LKR:MechReasEcon13}, we aim at making developers and users of mechanised reasoning systems, aware of \begin{enumerate*}\item new, challenging problems in the application domain of economics, of \item new target audiences not having the same background knowledge about formal languages, logics, etc., and thus of \item the necessity of enhancing the usability and documentation of the systems for a wider audience.\end{enumerate*}
Conversely, our message to economists, e.g.\ in a mechanised reasoning invited lecture at the 2012 summer school of the Initiative for Computational Economics (ICE~\cite{InitCompEcon}), is that there is a wide range of tools to assist with reliably solving relevant problems in economics.

\subsection{Infrastructure for the Community}
\label{sec:infrastructure}

With the \href{mailto:formare-discuss@cs.bham.ac.uk}{\texttt{ForMaRE-discuss@cs.bham.ac.uk}} mailing list and a project community site (both linked from our homepage), we furthermore provide infrastructure to the communities we intend to connect.  The main purpose of the community site is to collect pointers to existing formalisations of theorems, models and theories in economics~\cite{HundredTheoremsEconomics}, inspired by Wiedijk's list of formalisations of 100 well-known theorems~\cite{Wiedijk:100Theorems}, and to give a home to economics formalisations not published online otherwise.  The site is powered by Planetary~\cite{Kohlhase:ppte12}, a mathematics-aware web content management system with {\LaTeX} input, a format familiar to economists.

\subsection{Reaching out to Application Domains Beyond Economics}
\label{sec:reaching-out-beyond}

Finally, we are reaching out to further application domains beyond economics.  At our symposium on \emph{enabling domain experts to use formalised reasoning} (Do-Form~\cite{DoForm13}), economics and its formalisation was a strong showcase, with our expert collaborators working on auctions, matching and finance giving hands-on tutorials (cf.\ sec.~\ref{sec:establ-new-results}), but we also attracted submissions on domains as diverse as environmental models and autonomous systems and on tools from controlled natural language to formal specification.  Do-Form has aimed at connecting domain experts having problems (“nails”) and computer scientists developing systems (“hammers”) from the start of its novel submission and review process, which involved match-making.  We initially invited short hammer and nail descriptions.  We published the accepted submissions with editorial summaries and indications of possible matches\footnote{E.g., we pointed out to the authors of a hammer description that their system might be applicable to the problem mentioned in some nail description.}
online 
and then called for the second round of submissions: revisions of the initial submissions (now elaborating on possible matches), regular research papers or system descriptions, particularly encouraging new authors to match the initial submissions.  This finally resulted in 12 papers.

We believe that such community-building efforts, which originated from ForMaRE's goal to apply formal mathematical reasoning in economics, will also help to achieve closer collaboration \emph{within} the CICM community\footnote{This was one of the topics discussed in the 2012 MKM trustee election.}: In future, CICM attendees and reviewers reading this paper might point us to the best tools for formalising auctions, matching markets, and financial risk.

\printbibliography
\end{document}